\documentclass[conference]{IEEEtran}
\IEEEoverridecommandlockouts
\usepackage{cite}
\usepackage{amsmath,amssymb,amsfonts}
\usepackage{algorithmic}
\usepackage{graphicx}
\usepackage{textcomp}
\usepackage{xcolor}
\usepackage[bookmarks=false]{hyperref}
\usepackage{lineno,hyperref}
\usepackage{times}
\usepackage{latexsym}
\usepackage{multirow}
\usepackage{times}
\usepackage{xcolor}
\usepackage{soul}
\usepackage[utf8]{inputenc}
\usepackage[small]{caption}
\usepackage{graphicx} 
\usepackage{amsmath}
\usepackage{amsfonts}
\usepackage{mathrsfs}
\usepackage{bm}
\usepackage{array}
\usepackage{booktabs}
\usepackage{subfigure}
\usepackage{float}
\usepackage{url}
\usepackage{amsthm}
\usepackage[misc]{ifsym}
\usepackage[ruled,linesnumbered,vlined]{algorithm2e}

\theoremstyle{definition}
\newtheorem{definition}{Definition}

\newcommand{\model}{\textsc{RAU-GNN}}

\def\BibTeX{{\rm B\kern-.05em{\sc i\kern-.025em b}\kern-.08em
    T\kern-.1667em\lower.7ex\hbox{E}\kern-.125emX}}
\begin{document}

\title{Relevance-Aware Anomalous Users Detection in Social Network via Graph Neural Network}

\author{
\IEEEauthorblockN{Yangyang Li\IEEEauthorrefmark{1}\Letter, Yipeng Ji\IEEEauthorrefmark{2}, Shaoning Li\IEEEauthorrefmark{3}, Shulong He\IEEEauthorrefmark{4}, Yinhao Cao\IEEEauthorrefmark{3}, Yifeng Liu\IEEEauthorrefmark{1}, Hong Liu\IEEEauthorrefmark{5}\IEEEauthorrefmark{6}, \\ Xiong Li\IEEEauthorrefmark{7}, Jun Shi\IEEEauthorrefmark{8} and Yangchao Yang\IEEEauthorrefmark{8}}
\IEEEauthorblockA{
\IEEEauthorrefmark{1}National Engineering Laboratory for Public Safety Risk Perception and Control by Big Data, CAEIT, Beijing 100041, China;\\
\IEEEauthorrefmark{2}School of Computer Science and Engineering, Beihang University, Beijing 100191, China;\\
\IEEEauthorrefmark{3}School of Cyber Science and Technology, Beihang University, Beijing 100191, China;\\
\IEEEauthorrefmark{4}National Computer Network Emergency Response Technical Team/Coordination Center of China, Beijing 100029, China;\\
\IEEEauthorrefmark{5}School of Software Engineering, East China Normal University, Shanghai 200062, China;\\
\IEEEauthorrefmark{6} Shanghai Trusted Industrial Control Platform Co., Ltd., Shanghai 200062, China;\\
\IEEEauthorrefmark{7}Beijing Bitcat Technology Co., Ltd., Beijing, China;\\
\IEEEauthorrefmark{8}Shenzhen CyberAray Network Technology Co., Ltd, CETC, Shenzhen, China.\\
}
Email: 
\{liyangyang,liuyifeng3@cetc.com.cn\};
\{jiyipeng,19373215,caoyinhao\}@buaa.edu.cn; 
hsl@cert.org.cn;\\
liuhong@ticpsh.com;
li.xiong@foxmail.com;
jshi@nscslab.net;
forester@mail.ustc.edu.cn.
\IEEEcompsocitemizethanks{\IEEEcompsocthanksitem \Letter \quad Yangyang Li is the corresponding author}}

\maketitle

\begin{abstract}
Anomalous users detection in social network is an imperative task for security problems. Motivated by the great power of Graph Neural Networks(GNNs), many current researches adopt GNN-based detectors to reveal the anomalous users. However, the increasing scale of social activities, explosive growth of users and manifold technical disguise render the user detection a difficult task.
In this paper, we propose an innovate Relevance-aware Anomalous Users Detection model (\model) to obtain a fine-grained detection result. \model~ first extracts multiple relations of all types of users in social network, including both benign and anomalous users, and accordingly constructs the multiple user relation graph.
Secondly, we employ relevance-aware GNN framework to learn the hidden features of users, and discriminate the anomalous users after discriminating. 
Concretely, by integrating Graph Convolution Network(GCN) and Graph Attention Network(GAT), we design a GCN-based relation fusion layer to aggregate initial information from different relations, and a GAT-based embedding layer to obtain the high-level embeddings.
Lastly, we feed the learned representations to the following GNN layer in order to consolidate the node embedding by aggregating the final users' embeddings. We conduct extensive experiment on real-world datasets. The experimental results  show that our approach can achieve high accuracy for anomalous users detection. 
\end{abstract}


\section{Introduction}
Nowadays, social networks have become a non-substitutable platform for people's daily interactions and socialization. However, along with the large scale of online activities, some abnormal social users, enormous 'bots' accounts like zombie users, spammers, social bots, etc. have also shown up, which lurk in around benign users and have a non-negligible impact on the security of social networks\cite{nobata2016abusive}, IoT systems\cite{qian2020orchestrating}, IDC infrastructures\cite{hu2020toposch, wen2019ga,hase2016}, and the reliability of digital platforms\cite{tsc2017, yang2020performance}. Abnormal users are essential tools for orchestrated manipulation of mass media events and actively involved in the discussion of important events, including the public opinion direction of various political events\cite{yang2020rumor}. They are also responsible for disseminating less credible information or extreme ideologies, as well as increasing the level of confusion in some online discussions\cite{wang2019fdgars}. Therefore, detecting and filtering the anomalous users is essential for the security of social network\cite{gao2020hincti}.

Traditional approaches to anomalous users detection mainly concentrate on the explicit account information\cite{Beskow2019}(nickname, head portrait   etc.) collected from users' activities, and only treat users as isolated individuals. Furthermore, as a result of evolving techniques about generative adversarial networks(GANs)\cite{huang2020reinforced,wang2020seqst}, the anomalous users are able to imitate human-like behavior and disguise themselves. Hence the above static detection approaches are no longer adapted to the current situations, and the discrimination of anomalous users has become more ineffective\cite{zhu2020adversarial}. Recently, due to the perfect performance of graph neural networks(GNNs) in capturing the hidden connectivity in graph structure, many GNN-based anomaly detectors have been applied to various fraud or anomaly detection scenarios\cite{li2019spam, sun2020kollector}. In contrast to the traditional detection methods, GNN-based approaches consider the neighborhood information to learn the node representations with neural modules. They can be trained in an end-to-end diagram\cite{liu2020alleviating}, and their semi-supervised learning fashion dramatically decreases the labeling cost as well.

However, the existing GNN-based detection approaches still neglect the well-designed camouflage of the current newly anomalous users. They adjust their behaviors to alleviate the suspiciousness and would like to interact with normal users\cite{dou2020enhancing}. 
In other word, though users' external features and explicit relationships have been thoroughly exploited\cite{li2020modeling}, they still lead to the serious loss of detection accuracy in consequence of failing to distinguish the disguised relation between normal and anomalous users. As mentioned, anomalous users would build connection with some benign users to muddle through the feature-based detectors. Hence, extracting multiple relations from different users is of great significant to accurately classify the type of users most of which occupy semblable features\cite{zhao2021automatically, zhao2020multi}. Though some recent works have noticed the similar challenges, their solutions still can not fit with the anomalous users detection problems.

To further detect the well-disguised users in social network, in this paper, we propose a \textbf{GNN}-based \textbf{R}elevance-aware \textbf{A}nomalous \textbf{U}sers Detection (\textbf{\model}) model to achieve fine-grained anomalous users detection results.
We first extract the multiple relations from all users in social network. The relations between users could be roughly defined as an interaction, including retweets, comments or forwarding etc.\cite{peng2021streaming}. As for anomalous users, they prefer to forwarding similar blogs and take action at the same time. All these features can be constructed into a unified multiple relation graph.
Secondly, we leverage relevance-aware GNN-based framework to learn the hidden representation in the constructed relation graph from users. 
Concretely, we adopt GCN module to initially aggregate the structual information accross different relationships, and embedds the processed fusional features to the center nodes. 
Then we use multi-head GAT module to learn the high-level embeddings and we feed the final node embeddings to the following GNN layer, and aggregate all the users information from their neighbors, in order to consolidate the previous embeddings.  
Last we use the binary classification algorithm to classify the learned features, and discriminate the anomalous users.
We evaluate our proposed model with real-world datasets and the experimental results show that \model~can achieve high accuracy for anomalous users detection, outperforms other comparable baseline models. The main contributions of \model~are summarized as follows:
\begin{itemize}
    \item We extract different relations from users and accordingly establish a multiple relation users graph network as the basis of \model, and explore the importance of different users and relations.
    \item We propose a anomaly detecter named \model~based on a relevance-aware GNN framework, which consists of a GCN-based relation fusion layer, a GAT-based embedding layer and a final GNN aggregator respectively. The integration of these GNN layers can better learn the high-level representations and see through the well-designed disguise.
    \item Extensive experiments with real-world datasets are conducted to validate the effectiveness of \model~on anomalous users detection. The results demonstrate that our approach can achieve high accuracy and outperforms other classic comparable baseline models.
\end{itemize}
The rest of this paper is organized as follows. In Sec. \ref{sec:related}, we introduce the background and related work for anomalous users detection. Sec. \ref{sec:definition} demonstrates the necessary definition and models the detection problems. Sec. \ref{sec:framework} depicts the framework and components of our proposed \model~. We evaluate our proposed model with real-world datasets in Sec. \ref{sec:experiment} and analyze the experimental results with cases at length. Finally, we conclude this paper in Sec. \ref{sec:conclusion}.

\section{Related Work}\label{sec:related}
\subsection{Graph neural network}
GNNs\cite{scarselli2008graph} model is extended from the traditional neural network to implement graph-structured data. GNNs aim to learn  dimensional vector representations for nodes in the networks. The recent diagram of GNNs is to generalize convolutional operation(GCN)\cite{peng2019hierarchical}, which generates nodes' representation by aggregating their own and neighbors' features. GCNs address the cyclic mutual dependencies architectually and can be classified into spectral-based GCN and spatial-based GCN. Spectral-based GCNs\cite{peng2019fine} apply a normalized graph Laplacian matrix and graph Fourier transform to make spectral convolutional operations. Spatial-based GCNs directly convolve the nodes' representations with their close neighbors' to derive and propagate the updated representations. Compared with the spectral-based GCNs, spatial-based GNNs are more flexible and scalable with the sizes and the structures of the graphs. Furthermore, spatial-based GCNs are more efficient and are preferred in recent works\cite{peng2021lime}. Graph attention network (GAT)~\cite{GAT} introduces the classical attention  mechanism to GNNs. GAT outperforms in handling the long-term dependencies. It captures the content-based similarity between two entities and measures their distance of representations\cite{peng2019hierarchical,cao2021knowledge,peng2020motif}.

\subsection{Anomalous users detection}
Anomalous users detection mainly focuses on those who are mostly utilized as powerful tools to guide the direction of public opinion or carry out some crimes in social network\cite{du2017deeplog}. The most common anomalous users around are 1)\textit{Zombie user.} Zombie users are specifically fake followers who create an illusion of a high reputation and credibility via boosting the number of followers(similar to DDoS attack). Most of the time, zombie users remain silent without any interaction but suddenly show up  for some purpose. 2)\textit{Spammer.} Spammers aggressively post harmful content such as adult advertisements, e-magazines and links\cite{kaghazgaran2019wide}. Different from zombie users, spammers  interact with normal users by replying to comments, most of which are same in order to express the content and enhance their influence. 3)\textit{Social bots.} Social bots are designed to disseminate a certain point of view that can lead the direction of public opinion by posting content with certain hashtags\cite{hernandez2018social}. Social bots make disproportionate contribution in disseminating less credible information or extreme ideologies, as well as increasing the level of confusion on purpose.

To detect these anomalous users in social network, most approaches are proposed based on static personal features\cite{sun2020pairwise}. The basic Bayesian classification with user characteristics\cite{Cresci2015} and machine learning methods\cite{khalil2017} have proven efficient for zombie users detection. Recently, many GNN-based detectors show better performance than traditional methods. Wang et al.\cite{wang2019fdgars} leverage graph convolutional network  for fraudster detection in the online app review system. Li\cite{li2019spam} et al. design a GCN-based anti-Spam model which integrates a heterogeneous graph and a homogeneous graph to capture the local context and global context of a comment, and aggregate neighbor's information. Zhang et al.\cite{zhang2019key} leverage convolution mechanism to learn embeddings of each single-view attributed graph and attention mechanism to fuse different embeddings.

\section{Problem Definition and Modeling}\label{sec:definition}
In this section, we will first give the necessary notations and definition of this paper in Table \ref{table:t1}, and model the anomalous users detection problems.

\begin{definition}\textbf{Multiple Relation Graph.}\label{def:mrg}
A multiple relation graph is $\mathcal{G}$ that contains a set of relation graph $U^r = \{(\mathcal{V},\{\mathcal{E}_r\}_{r=1}^R), Y\}$, where $\mathcal{V}$ stands for the collections of user nodes $\{v_1, v_2, \cdots, v_n\}$, and $\mathcal{E}_r$ stands for edges $e_{i,j}^r = (v_i, v_j)$ under relation $r \in R$. $R$ is the total number of relations. $Y$ is the node label.
\end{definition}
\begin{figure}[htbp]
    \centering
    \includegraphics[width=.45\textwidth]{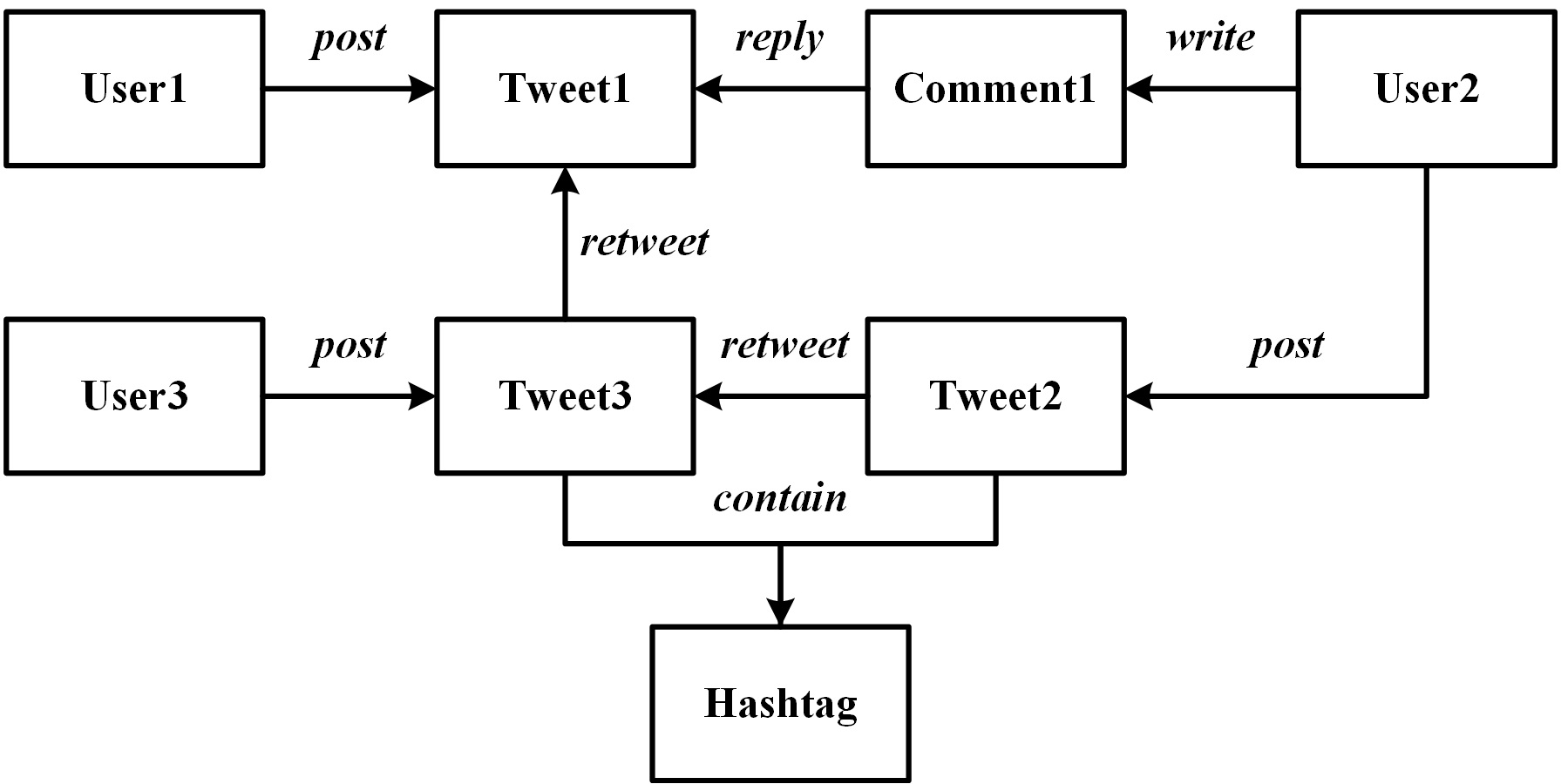}
    \caption{An example of users interaction in multiple relation graph}
    \label{fig:HIN}
\end{figure}

Fig.\ref{fig:HIN} gives an example of multiple relation network for users, including various relationships between users. Generally, users' interactions in social network can be categorized into \textit{following, commenting/replying, posting/retweeting}, and \textit{hashtagging}, which also represent different types of relation between nodes in constructed multiple relation graph. We denote these operations as $\{f, c, p, h\}$, respectively, which stand for the different types of relations in constructed multiple relation graph.

\begin{definition}\textbf{Anomalous Users Detection on Graph.}
For the anomalous users detection problem, we aim to justify whether the given node $v_i$ is abnormal. Hence we can treat the discrimination problem as a binary classification on the multiple relation graph. The label of nodes in multiple relation graph is $y_v \in \{0, 1\}$, where label $0$ represents the benign user and label $1$ represents the anomalous user that to be detected. The graph-based anomalous user detection algorithm learns the detection model $f({V}_i, {E}_r; \theta)$ based on the labeled user nodes' information under each relations, and predict whether the unlabeled nodes in graph are anomalous users. $\theta$ denotes the parameters in model. 

\end{definition}

\begin{definition}\textbf{GNN-based User Detection.}
A Graph Neural Network(GNN) is a layer-wise deep learning framework, aiming to embed and learn graph features by aggregating information from its neighbor nodes. Here we give the unified formulation of GNNs from perspective of neighbor aggregation to give a full picture of proposed GNN-based framework:
\begin{equation}
    h^{(l+1)}_{v_i} = \sigma\left(\underset{\forall v_j \in N(v_i)}{AGG} (h^{(l)}_{v_i}\oplus h^{(l+1))}_{v_j}   \right).
\end{equation}
For the center node $v_i$, $h^{(l+1)}_{v_i}$ is its hidden representation at the $(l+1)$-th layer, and we define $h^{(0)}_{v_i} = x_i$ is the input feature. $h^{(l)}_{v_i}$ is the input of $(l+1)$-th layer in GNN, that is to say, GNNs follow the propagation rule. $AGG$ is the aggregation function such as mean aggregation or attention aggregation. $\oplus$ is the fusion operation (concatenation or summation) that combine the extracted features from its neighbor. The final purpose of GNN-based user detection is to make use of the advantage of GNNs on aggregating the information of neighborhood. The node embeddings at the last layer of GNN represent the kernel of each users, and we use these high-level embeddings to discriminate whether the users are anomalous.
\end{definition}

\renewcommand\arraystretch{1.5}
\begin{table}[t]
    \caption{Glossary of Notations.}
    \resizebox{\linewidth}{!}{%
    \begin{tabular}{r|l}  
    \toprule
      \textbf{Notation} & \textbf{Description}\\
      \midrule
      $r;R$ & Relation; Total number of relations \\
      $\mathcal{G}; U^r$ & Multiple relation graph; Homo user graph under relation $r$  \\ 
      $\mathcal{V};\mathbf{E}; X, $ & Node set; Edge set; Features \\
      $W$ & Sub-matrix of heterogeneous graph \\
      $\Tilde{D}$ & Diagonal matrix \\
      $\mathbf{H}^{(l+1)}_r$ & Embedding at layer $(l+1)$ under relation $r$ in GCN module\\
      $z_{v_i}$ & Normalized Embedding for node $v_i$ \\
      $\oplus$ & Fusion operation \\
      $\operatorname{norm}$ & Normalization \\
      $\mathbf{h}^{(l+1)}_r$ & Embedding at layer $(l+1)$ under relation $r$ in GAT module\\
      $AGG$ & Aggregation operation \\
      $\mathcal{Z}^{(l+1)}$ &  Embedding at layer $(l+1)$ in enhanced aggregator\\
      $\mathcal{Z}_v$ & Final representation for user $v$\\
      $\mathcal{P}_v$ & Possibility for node $v$ to be an anomalous user\\
      $\mathcal{L}_{\model}$ & Loss function for \model~model\\
      \bottomrule
    \end{tabular}}\label{table:t1}
\end{table}

\section{Relevance-Aware Anomalous Users Detection}\label{sec:framework}

\begin{figure*}
    \centering
    \includegraphics[width=1.\textwidth]{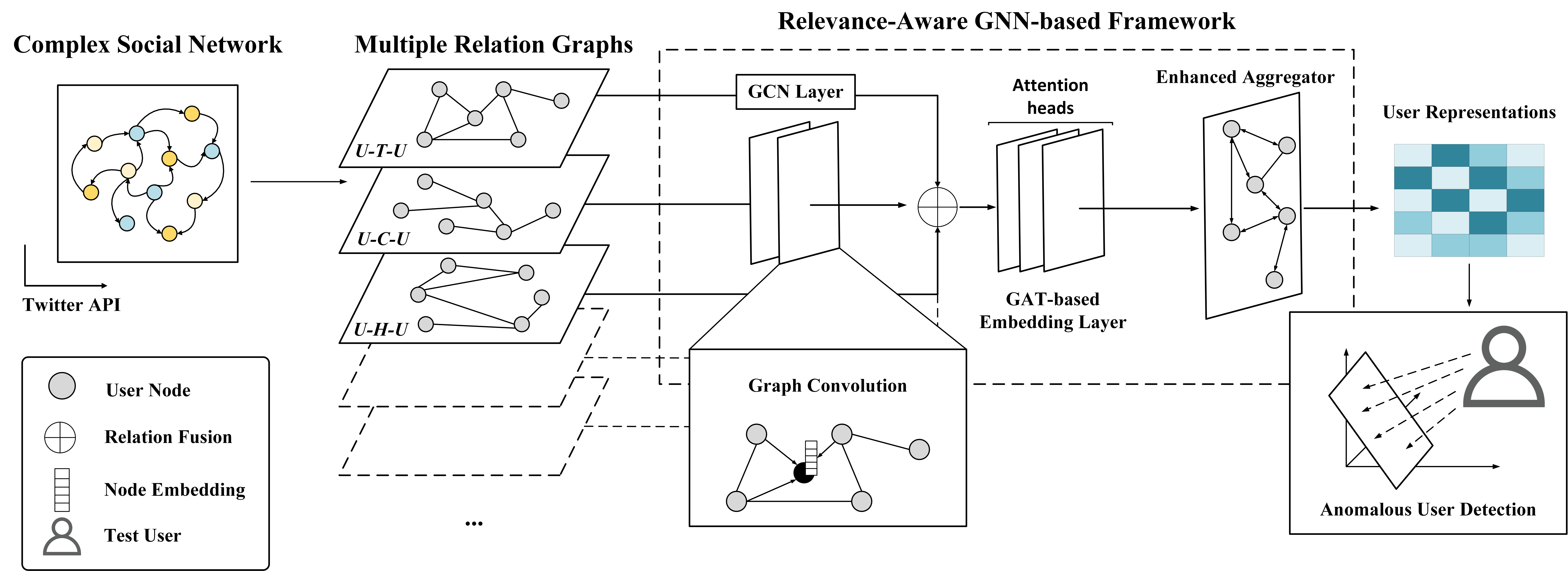}
    \caption{Relevance-Aware Anomalous Users Detection}
    \label{fig:framework}
\end{figure*}

\subsection{Overall Framework}
The overall framework of \model~proposed in this paper is shown in Fig.\ref{fig:framework}. 
Concretely, our \model~contains three processes, including the construction of multiple relation graph for users, an integrating GNN-based framework to learn the high-level representations, and a discrimination layer to detect the anomalous users. The proposed GNN-based framework integrates three layers, including a GCN-based relation fusion layer, a GAT-based embedding layer and a final GNN aggregator.
The following sections will introduce more details of \model~at length.
\subsection{Construction of Multiple Relation Graph}
\label{sec:mrg}
To begin with, we extract the features of users from \textit{user} object by Twitter API\footnote{https://developer.twitter.com/en/docs/twitter-api}. The rich extracted features serve as the initial node features in social network. 
During the preprocessing, we aim to unify all the users and their relations, and break them down to a set of relation-based homogeneous graphs.
First, we extract the user-oriented elements, such as tweets, comments etc. and organize them in a unified manner to represent the different relations. 
For example, we extract a tweet that reveals User1 comments on tweet of other users. We denotes User1 as $v_1$ and add edge  between $v_1$ and the extracted tweet comments. We repeat the same process for all the users in social network, and obtain a complex heterogeneous graph.

Next, we break down the initial graph to multiple relation graphs. The multiple relation graph only contains user nodes and with different edge attributes representing different types of relations. Specifically, the separating process is defined as follows:
\begin{equation}
    U^r_{i, j} = \left[ \sum_r W_{vr} \cdot W_{vr}^T \right], r \in \{f, c, p, h\}.
\end{equation}
Here adjacency matrix $U^r_{N \times N}$ is the homogeneous user graph under relation $r$, where $N$ is the total number of users in graph. $\cdot_{i.j}$ denotes the matrix element at the $i^{th}$ row and $j^{th}$ column, and $r$ denotes the element type under the mentioned operations.  $W_{vr}$ is a sub-matrix denotes rows of user nodes and columns of type $r$. $\cdot$ is matrix transpose.

Lastly we give the extracted user feature vectors to the multiple relation graphs. The initial feature is denoted as $X = \{x_v \in R^d\}$, where $x_v$ is the initial feature of user with $d$ dimension.  We collect all types of $\{U^r\}|^R_{r=1}$ and obtain the processed multiple relation graphs $\mathcal{G} = (X, \{U^r\}|^R_{r=1}, Y)$.

\subsection{Relevance-Aware GNN-based Framework}
The proposed GNN-based framework consists of a GCN-based relation fusion layer, a GAT-based embedding layer and a final GNN aggregator, respectively.
\subsubsection{GCN-based Relation Fusion Layer}
We implement GCN module to learn the discriminating user node representation in each relation, and fuse the learned features in different relations to the center node. The layer-wise propagation rule of multi-layer GCN is defined as follows:
\begin{equation}
\label{equ:gcn}
    \mathbf{H}^{(l+1)}_r = \sigma \left(\Tilde{D}^{-\frac{1}{2}}\Tilde{U}_r\Tilde{D}^{-\frac{1}{2}}H_r^{(l)}W^{(l)} \right),
\end{equation}
where $\Tilde{U}_r = U_r + I_N$, $r$ denotes the relation type of graph $U$ and $N$ denotes the total number of users in this graph. $\Tilde{D}$ is diagonal matrix with $\Tilde{D}_{ii}=\sum_j\Tilde{U}_{ij}$, $I_N$ is the identity matrix, $W$ is the parameter matrix and $l$ is the number of layers. $\sigma$ denotes the activation function, e.g., Sigmoid or ReLU. The input layer of GCN module is $H^{(0)} = X$, and $X$ is the initial features constructed in Section \ref{sec:mrg}. We use $H^r_{v_i}$ as the final representation of user node $v_i$ under relation $r$.

Next we fuse the output representations from GCN layers in different relations, and obtain the final dense representations of the center node.
We give the fusion formulation as follows:
\begin{equation}
\label{equ:fusion}
    z_{v_i} = \operatorname{norm}\left(H^{(r=1)}_{v_i} \oplus H^{(r=2)}_{v_i} \oplus \cdots \oplus H^{R}_{v_i} \right),
\end{equation}
where $v_i$ is the center node. $H^{(r)}_{v_i}$ is the output representation under relation $r \in R$ with $d'$ dimension. $\oplus$ denotes the operator that combines information of $v_i$ across all its all relations, e.g., concatenation or summation. $\operatorname{norm}$ denotes the normalization process to avoid overfitting.
It is noting that after the GCN-based fusion process, we unify the multiple relation graphs to one complete homogeneous user graph $\mathcal{A}$ with the combined representations, wherein all the users are involved. And we implement this unified graph to facilitate further processing.

\subsubsection{GAT-based embedding layer}
In order to prick camouflage of anomalous users, we leverage GAT module to  study the correlations between users in social network, and embed the fusion features in user graph. To be specific, we define the GAT layer-wise propagation rule as follows:
\begin{equation}
\label{equ:gat}
    \textbf{h}_{v_i}^{(l+1)} \leftarrow  \overset{heads}{\Vert} \left(\textbf{h}_{v_i}^{(l)} \oplus \underset{\forall v_j \in N(v_i)}{AGG}(\textbf{h}_{v_j}^{(l)}) \right).
\end{equation}
Here $\textbf{h}_{v_i}^{(l+1)}$ is the representation of user $v_i$ at the $(l+1)$th GAT layer, and the input $\textbf{h}_{v_i}^{(0)}$ is the output of the fusion process $z_{v_i}$.
$N(v_i)$ stands for a set of neighbors of user $v_i$ according to graph $\mathcal{A}$. $\oplus$ denotes the operator that combines information of $v_i$ with its neighbors. $\overset{heads}{\Vert}$ denotes head-wise concatenation. $AGG$ represents the aggregation function that mapping neighborhood information into a vector. Here we adopt the attention aggregation. We use $\textbf{h}_{v_i}$ as the final representation with dimension $d'$ after GAT-based user embedding.

The learned representation are detector's cognition about the users in social network, which are also a fusion of natural language semantics, temporal activities, and the structural information of the completed user graph. These high-level representations stand for the kernel of a user, and helps to identify the abnormal account.

\subsubsection{Enhanced GNN Aggregator}
Based on the relation fusion layer and embedding layer, we have obtained the hidden representations of each users in social network. In order to achieve a fine-grained detection result and be able to adaptive to the real-world scenario, we add a following enhanced GNN aggregator to consolidate the previous embeddings. We define the GNN aggregation as follows:
\begin{equation}
\label{equ:agg}
    \mathcal{Z}^{(l+1)} = \sigma\left(\underset{\forall v_j \in N(v_i)}{AGG} (\mathcal{Z}^{(l)}_{v_i}\oplus \mathcal{Z}^{(l+1))}_{v_j}   \right),
\end{equation}
where $\mathcal{Z}^{(l)}_{v_i}$ is the center node embedding at the $l$-th layer, $\mathcal{Z}^{(l+1))}_{v_j}$ is the neighbor node embedding at the $(l+1)$-th layer. 
$\oplus$ denotes the embedding summation. $AGG$ represents the mean aggregation. $\sigma$ denotes the activation function. Here we adopt ReLU non-linear function.

For the GNN aggregator is mostly applied to consolidate the previous learned embeddings instead of learning more high-level semantics, we fix the output dimension of GNN aggregator as $d'$, which is the same as the output dimension of GAT-based embedding layer, and only adopt one layer GNN. 


\subsection{Optimization}
After the enhanced GNN aggregator, for each center user node $v$, its final embedding is the output of aggregator $\mathcal{Z}_{v}$. Therefore, we accordingly design the loss of \model~as a cross-entropy loss function to minimize the ground-truth and predicted label, and achieve the anomalous detection:
\begin{equation}
\label{equ:loss}
    \mathcal{L}_{\operatorname{\model}} = -\underset{v\in\mathcal{V}}{\sum} \log (y_v \cdot \mathcal{P}_v) + \lambda \Vert \Theta \Vert_2,
\end{equation}
where $\mathcal{V}$ is all collection of node in user graph, $y_v$ denotes the label of node $v$, and $\mathcal{P}_v$ denotes the probability of \model's prediction. We define $\mathcal{P}_v = \sigma\left(\operatorname{MLP}(\mathcal{Z}_v)\right)$. $\sigma$ is the activation function. We adopt Sigmoid in \model. $\operatorname{MLP}$ denotes the multi-layer perception. $\lambda$ is the weight parameter and we add $\Vert \Theta \Vert_2$, which represents the $L_2$-norm of all module parameters in GNN-based framework, to obtain the well generalization.

\begin{algorithm}[h]
\SetAlgoVlined
\KwIn{A set of Multiple Relation Graphs with nodes features and labels: $\mathcal{G} = (X, \{U^r\}|^R_{r=1}, Y)$,
Number of Layers, Mini-batches: $L, B$.}
\KwOut{User Representation $\mathcal{Z}_{v}$, $\forall v \in \mathcal{V}_{train}$.}
$\mathbf{H}^{(0)}_v \gets x_v$\;

\For(\tcp*[h]{Train in mini-batches}){$b = 1, 2, ..., B$}{
\For{$r=1,2,\cdots,R$}{
    \For{$l=1,2,\cdots,L$}{
        $\mathbf{H}_v^{(l+1)} \gets$ Eq. (\ref{equ:gcn})\;
    }
    }
    $\mathbf{H}_v^{r} \gets \mathbf{H}_v^{(L)}$\;
$z_v \gets$ Eq. (\ref{equ:fusion})\;
$\mathbf{h}_v^{(0)} \gets z_v$\;
\For{$l=1,2,\cdots,L$}{
    $\mathbf{h}_v^{(l+1)} \gets$ Eq.(\ref{equ:gat})\;
}
$\mathbf{h}_v \gets \mathbf{h}_v^{(L)}$\;
$\mathcal{Z}^{(0)}_v \gets \mathbf{h}_v$\;
$\mathcal{Z}_v \gets$ Eq. (\ref{equ:agg})\;
$\mathcal{L}_{\operatorname{\model}} \gets$ Eq. (\ref{equ:loss})\;
Back-propagation to update parameters\;
}

\caption{\model: GNN-based Relevance-Aware Anomalous Users Detection}
\end{algorithm}

\section{Experiments and Evaluation}\label{sec:experiment}
\subsection{Dataset and Graph Construction}
We utilize the popular YelpChi review dataset, along with twitter datasets to study the anomalous user detection problems.  
The YelpChi dataset includes hotel and restaurant reviews filtered (spam) and recommended (legitimate) by Yelp. 
In Twitter dataset, we manually label users with more than 20\% helpful entities from anomalous users.
In this paper, we treat the spammers as anomalous users, and conduct the anomalous detection task on YelpChi and Twitter dataset, which can also be considered as a binary classification problem.

Similar to definition mentioned in Def. \ref{def:mrg}, we extract the reviews, products and time. 
In Twitter, we take tweets, comments and hashtags instead. We then construct the multiple relation graph for YelpChi and Twitter dataset. The details of construction is demonstracted in Section \ref{sec:mrg}. The statistics are shown in Table \ref{table:stc} with some explanations:

\renewcommand\arraystretch{1.5}
\begin{table}[htbp]
\caption{Statistics in Dataset and Graph}
\label{table:stc}
\centering
\begin{tabular}{llcccllll}
\toprule
                                                       & \multicolumn{1}{c}{\textbf{Nodes}} & \textbf{Anomalous\%}    & \textbf{Relation}    & \textbf{Edges}  \\ \cline{1-5}
\multicolumn{1}{c|}{\multirow{4}{*}{\textbf{YelpChi}}} & \multirow{4}{*}{50,128}            & \multirow{4}{*}{15.1\%} & R-U-R                & 51,715        \\
\multicolumn{1}{c|}{}                                  &                                    &                         & R-P-R                & 582,462          \\
\multicolumn{1}{c|}{}                                  &                                    &                         & R-T-R                & 4,402,892       \\
\multicolumn{1}{c|}{}                                  &                                    &                         & ALL                  & 5,037,069        \\ \midrule
\multicolumn{1}{l|}{\multirow{4}{*}{\textbf{Twitter}}} & \multirow{4}{*}{12,384}            & \multirow{4}{*}{24.3\%} & U-T-U                & 2,625,142       \\
\multicolumn{1}{l|}{}                                  &                                    &                         & U-C-U                & 516,706  \\
\multicolumn{1}{l|}{}                                  &                                    &                         & U-H-U                & 311,610      \\
\multicolumn{1}{l|}{}                                  &                                    &                         & ALL                  & 4,206,916    \\ \bottomrule
\end{tabular}
\end{table}
\textbf{YelpChi}.
1) \textit{R-U-R}: connects reviews sent by same users.
2) \textit{R-P-R}: connects reviews under same product.
3) \textit{R-T-R}:connects reviews that created in the same month.
\textbf{Twitter}.
1) \textit{U-T-U}: connects users that mention the same tweet.
2) \textit{U-C-U}: connects users that comments the same content.
3) \textit{U-H-U}: connects users that involved in same hashtag.
The number of edges belonging to each relations is also shown in Table \ref{table:stc}.

\subsection{Baseline Algorithm}
To verify the effectiveness of \model, we compare our proposed model against baselines, including general GNN models and new GNN-based methods. Details of the selected baseline algorithm are demonstrated as follows:

\textit{GCN} and \textit{GAT} are  semi-supervised homogeneous graph models that use convolution or attention mechanism to aggregate neighborhood information of graph nodes.
\textit{HGT}\cite{hu2020heterogeneous} characterize the heterogeneous attention over different types of nodes, and it performs cross-layer messages from different types of neighbors for higher-order aggregation.
\textit{GraphSAGE} is an inductive framework that leverages node attribute information to efficiently generate representations on previously unseen data.
\textit{FDStar}\cite{wang2019fdgars} is a graph convolutional network approach for fraudster detection in review system.
\textit{GAS}\cite{li2019spam} is a GCN-based Anti-Spam model, capturing the local and global context of a comment to detect spammers.
\textit{SemiGNN} is a semisupervised attentive graph neural network that utilizes the multi-view labeled and unlabeled data for fraud detection. 

Furthermore, to inspect the validation of relevance-aware GNN-based framework, we decompose the \model~into a plain relation \model(PR-\model), which directly sum up the initial embedding across relations without GCN module, and a plain aggregation \model(PA-\model), which removes the last enhanced GNN aggregator. 

\subsection{Experimental Setting}
Due to the small percentage of anomalous users in dataset, we adopt mini-batch training techinique to efficiently train \model.
During the detection process, we set the output dimension of final embeddings 64, batch size 1024 for YelpChi and 256 for Twitter, learning rate 0.005, $L_2$ regularization weight $\lambda$ 0.001.
The parameters of \model~are initialized with Xavier and we adopt Adam optimizer. 
All experiments are conducted on a 64 core Intel Xeon CPU E5-2680 v4\@2.40GHz with 256GB RAM and 1$\times$NVIDIA Tesla P100-PICE GPU. We implement \model~and other comparison models with Python 3.7.1 and Pytorh 1.6.0.

\begin{table*}[htbp]
\centering
\caption{Anomalous detection performance (\%) on two datasets under different percentage of training data.}
\label{table:all}
\begin{tabular}{|cccccccccc|ccc|}
\toprule
                                                        & \textbf{Metric}                                & \textbf{Train\%}          & \textbf{GCN} & \textbf{GAT} & \textbf{HGT} & \textbf{\begin{tabular}[c]{@{}c@{}}Graph-\\ Sage\end{tabular}} & \textbf{GAS} & \textbf{FDStar} & \textbf{Semi-GNN} & \textbf{PR-RAU} & \textbf{PA-RAU} & \textbf{RAU-GNN} \\ \hline
\multicolumn{1}{|c|}{\multirow{8}{*}{\textbf{YELP}}}    & \multicolumn{1}{c|}{\multirow{4}{*}{Accuracy}} & \multicolumn{1}{c|}{10\%} & 52.12        & 54.06        & 58.83        & 53.52                                                          & 61.54           & 66.82        & 52.12             & 55.63           & 69.16           & \textbf{70.04}   \\
\multicolumn{1}{|c|}{}                                  & \multicolumn{1}{c|}{}                          & \multicolumn{1}{c|}{20\%} & 53.21        & 57.43        & 60.04        & 54.37                                                          & 62.85           & 65.76        & 51.86             & 56.14           & 70.03           & \textbf{70.62}   \\
\multicolumn{1}{|c|}{}                                  & \multicolumn{1}{c|}{}                          & \multicolumn{1}{c|}{30\%} & 53.41        & 57.26        & 59.43        & 55.08                                                          & 61.28           & 65.43        & 51.65             & 55.82           & 70.42           & \textbf{70.96}   \\
\multicolumn{1}{|c|}{}                                  & \multicolumn{1}{c|}{}                          & \multicolumn{1}{c|}{40\%} & 52.34        & 56.48        & 58.61        & 54.24                                                          & 62.04           & 65.27        & 51.69             & 55.75           & 70.94           & \textbf{71.20}   \\ \cline{2-13} 
\multicolumn{1}{|c|}{}                                  & \multicolumn{1}{c|}{\multirow{4}{*}{Recall}}   & \multicolumn{1}{c|}{10\%} & 51.32        & 53.45        & 56.12        & 52.35                                                          & 56.10           & 54.33        & 52.37             & 54.41           & 63.51           & \textbf{63.60}   \\
\multicolumn{1}{|c|}{}                                  & \multicolumn{1}{c|}{}                          & \multicolumn{1}{c|}{20\%} & 53.41        & 53.95        & 56.13        & 52.84                                                          & 55.62           & 55.14        & 52.41             & 54.36           & 64.82           & \textbf{64.78}   \\
\multicolumn{1}{|c|}{}                                  & \multicolumn{1}{c|}{}                          & \multicolumn{1}{c|}{30\%} & 52.76        & 54.11        & 55.84        & 52.63                                                          & 55.68           & 54.73        & 52.18             & 54.69           & 65.44           & \textbf{65.52}   \\
\multicolumn{1}{|c|}{}                                  & \multicolumn{1}{c|}{}                          & \multicolumn{1}{c|}{40\%} & 51.92        & 54.36        & 55.92        & 52.82                                                          & 55.71           & 54.58        & 51.59             & 54.58           & 65.43           & \textbf{65.73}   \\ \midrule\toprule
\multicolumn{1}{|c|}{\multirow{8}{*}{\textbf{Twitter}}} & \multicolumn{1}{c|}{\multirow{4}{*}{Accurary}} & \multicolumn{1}{c|}{10\%} & 68.54        & 67.72        & 71.57        & 65.86                                                          & 75.41           & 72.18        & 68.20             & 69.12           & 78.38           & \textbf{78.42}   \\
\multicolumn{1}{|c|}{}                                  & \multicolumn{1}{c|}{}                          & \multicolumn{1}{c|}{20\%} & 69.23        & 68.57        & 70.83        & 66.93                                                          & 76.27           & 73.45        & 66.53             & 69.75           & 78.92           & \textbf{79.71}   \\
\multicolumn{1}{|c|}{}                                  & \multicolumn{1}{c|}{}                          & \multicolumn{1}{c|}{30\%} & 69.14        & 67.21        & 71.46        & 67.05                                                          & 73.56           & 74.37        & 65.94             & 70.11           & 79.58           & \textbf{79.82}   \\
\multicolumn{1}{|c|}{}                                  & \multicolumn{1}{c|}{}                          & \multicolumn{1}{c|}{40\%} & 68.43        & 68.13        & 70.92        & 67.51                                                          & 72.32           & 74.65        & 66.82             & 69.67           & 79.26           & \textbf{79.64}   \\ \cline{2-13} 
\multicolumn{1}{|c|}{}                                  & \multicolumn{1}{c|}{\multirow{4}{*}{Recall}}   & \multicolumn{1}{c|}{10\%} & 60.42        & 61.24        & 65.16        & 64.73                                                          & 68.31           & 69.55        & 63.35             & 65.13           & 70.49           & \textbf{70.53}   \\
\multicolumn{1}{|c|}{}                                  & \multicolumn{1}{c|}{}                          & \multicolumn{1}{c|}{20\%} & 59.03        & 63.41        & 66.69        & 64.65                                                          & 68.65           & 69.64        & 64.43             & 66.29           & 70.38           & \textbf{71.49}   \\
\multicolumn{1}{|c|}{}                                  & \multicolumn{1}{c|}{}                          & \multicolumn{1}{c|}{30\%} & 59.75        & 62.87        & 65.86        & 65.69                                                          & 69.29           & 68.82        & 63.87             & 66.08           & 71.06           & \textbf{71.84}   \\
\multicolumn{1}{|c|}{}                                  & \multicolumn{1}{c|}{}                          & \multicolumn{1}{c|}{40\%} & 59.58        & 62.13        & 66.73        & 66.56                                                          & 68.45           & 68.57        & 63.31             & 65.74           & 71.35           & \textbf{71.73}   \\ \bottomrule
\end{tabular}
\end{table*}

\section{Overall Evaluation}

In this section, we demonstrate the experimental results in detection effectiveness, accuracy and the impact of parameters of \model.

We adopt Accuracy score and Recall to quantitatively evaluate the effectiveness of detection. In our experiments, the performances are reported with the best results.
We choose different percent of the data samples for training \model~and the residual are organized for testing. Table~\ref{table:all} illustrates the Accuracy scores and Recall of each model. It is obvious that \model~outperforms all other baselines and obtains better classification accuracy. Generally,  models adapted for heterogeneous graph perform better compared with other homograph-based models. 
In summary, \model~achieves dramatic improvement in the accuracy and effectiveness of anomalous users detection.
We evaluate the performance of models at length in the following sections.

\subsection{Single Relation and Multiple Relation}
As for all the baseline models in Table \ref{table:all}, GCN, GAT, GraphSage and PR-\model~are implemented on single relation graph that all relations are merged together, wherein PR-\model~runs on the simple combination of multiple relation graph.
HGT, FDStar, GAS and Semi-GNN are implemented on the constructed multiple relation graph. Compared with the performances of single relation models, GNN-based model on multiple relation graph generally obtain better results on accuracy score and recall. Among all the multi-relation GNNs baselines, GAS outperforms all other models, for GAS additionally consider the local contents to enhance the classifier. Better than GAS, \model~aggregates information from different relations, and consolidate the node embedding on a homogeneous user graph. The experimental results show that \model~can better filter the anomalous users in social network. It also demonstrates the significance of relations between users when there are more anomalous users lucks around, and verifies the behavior preference of anomalous users that widely connects with benign users. It is noting that the performance of PR-\model~is similar to other single relation GNN model, which proves the effectiveness of multiple relation aggregation.

\subsection{Training Percentage Analysis}
From Table \ref{table:all}, we can find there are little difference among the training percentages. Even if the training percentage is increasing, the fluctuation of accuracy score and recall is small, and maintain at a certain level. The experimental results of gapped training percentages demonstrate the advantage of semi-supervised learning, where a small amount of labeled nodes is enough for training a model, and would achieve better classification results.

\begin{figure}[htbp]
    \centering
    \includegraphics[width=.5\textwidth]{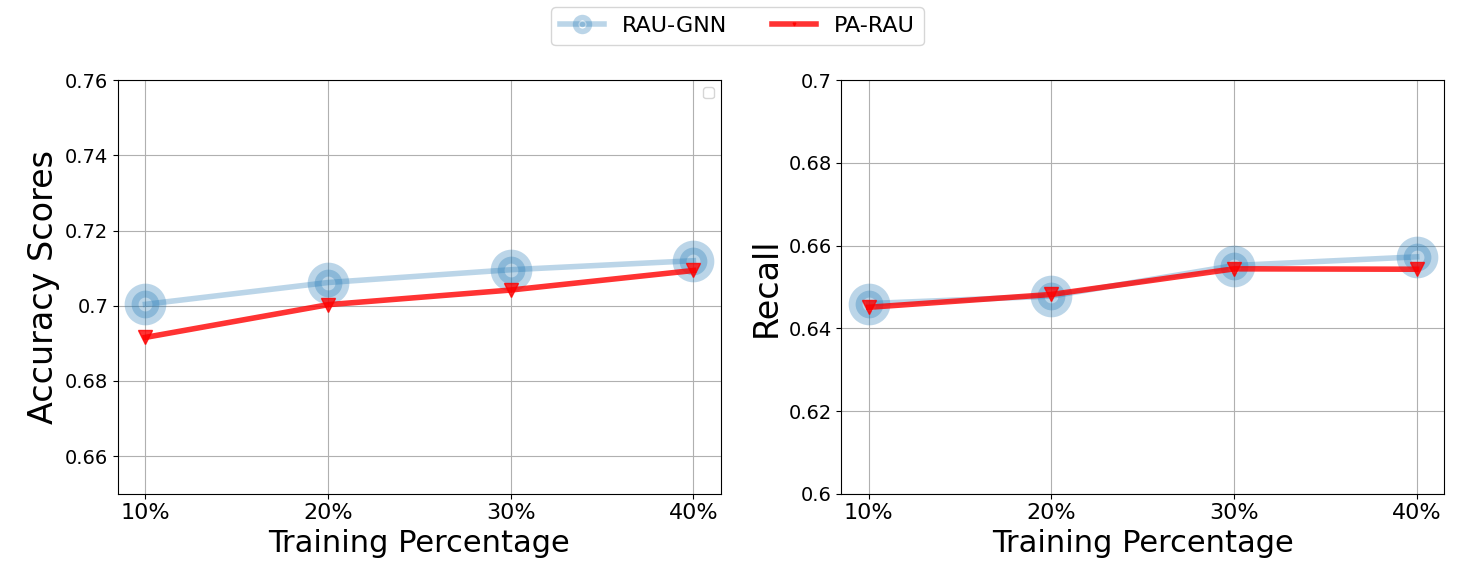}
    \caption{Model Performance under Different Training Percentage}
    \label{fig:my_label}
\end{figure}

\subsection{Variants of \model}
As mentioned above, we decompose the \model~into two plain model variants.
PA-\model~directly sum up the initial embedding across relations without GCN module.
We use PA-\model~to verify the advantage of aggregation in multiple relation.
PA-\model removes the last enhanced GNN aggregator, and use the output from GAT module as the final embeddings. The last 3 columns in Table \ref{table:all} shows the performance of \model~and proposed variants.
It is obvious that \model~outperforms than all other baselines.
The performance of PR-\model~is mentioned in multiple relation analysis. It proves the feasibility of cross-relation aggregation. PA-\model~achieves better performance, for it retains the GCN-based relation fusion layer. The improvement of \model~compared with PA-\model~verifies the effect of enhanced aggregator. Due to that the last aggregator is used to improve the robustness and generalization of model, it only brings little promotion on accuracy score and recall.

\subsection{Hyperparameter Sensitivity}
\begin{figure}[htbp]
    \centering
    \includegraphics[width=.5\textwidth]{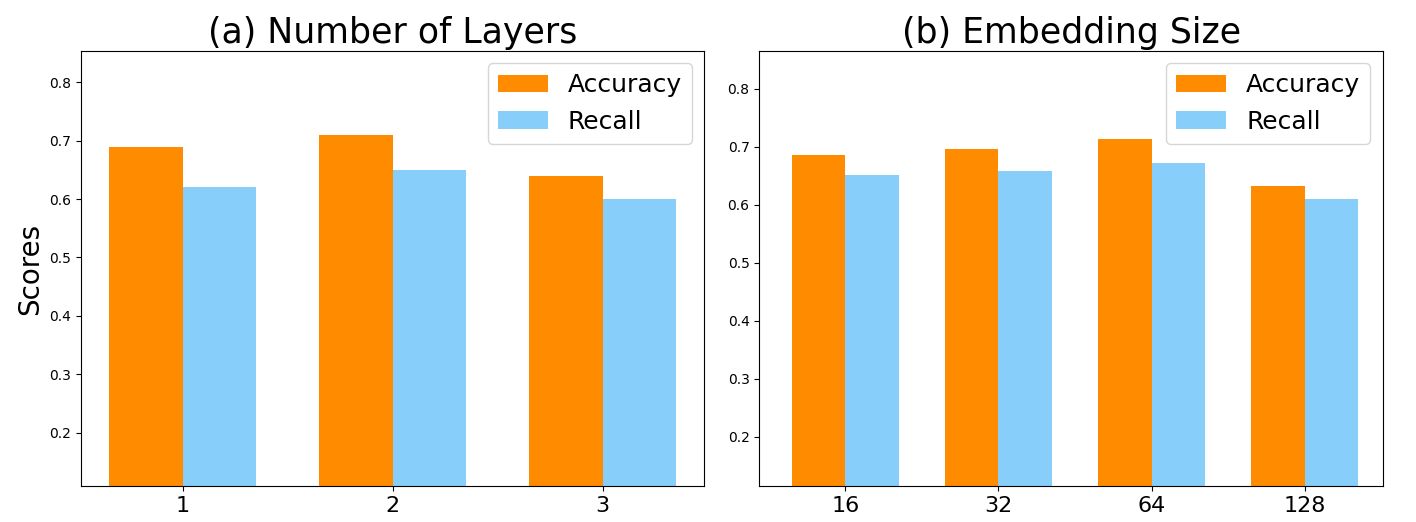}
    \caption{Hyperparameter Sensitivity Analysis}
    \label{fig:hyper}
\end{figure}

We analyze the hyperparameter in this subsection. Figure \ref{fig:hyper} shows the testing performance of \model. To analyze the number of layers in GCN-based fusion layer, we observe the results by increasing the number of layers. Figure \ref{fig:hyper}(a) shows the performance of different layer numbers on YelpChi dataset. We can see a peak when the number of layer is set to 2. For the 3-layer GCN, \model~confronts with the overfitting problem, and obtain a worse result.
Figure \ref{fig:hyper}(b) shows the effect of different embedding size. We set the output embedding size to 16, 32, 64 and 128. It is obvious that embedding szie with 64 would achieve better results.

\section{Conclusion}\label{sec:conclusion}
In this paper, we develop a new GNN-based relevance-aware anomalous user detection model, named \model, to effectively discriminate the wel-disguised anomalous users in social network.
Firstly, \model extracts multiple relations between users in social network, and accordingly constructs the multiple user relation graph.
Secondly, we design relevance-aware GNN framework to learn the high-level of users, and discriminate the anomalous users through discriminating. 
Concretely, we design a GCN-based relation fusion layer to aggregate initial information from different relations through convolutional operation, and a GAT-based embedding layer to represent the hidden embeddings of users.
Lastly, we feed the learned representations to the following GNN aggregator in order to get the node embedding by aggregating the final users' embeddings, and develop the robustness and generalization of \model.   The experimental results  show that our approach can achieve better accuracy for anomalous users detection. 

\section*{Acknowledgment}
The authors of this paper were supported by NSFC through grant U20B2053, and S\&T Program of Hebei through grant 20310101D,
Major Science and Technology Plan Projects of Hainan through grant ZDKJ2019008,
This work was also supported by the Opening Project of Shanghai Trusted Industrial Control Platform.

\bibliographystyle{IEEEtran}

\bibliography{mybib}

\end{document}